\documentclass[preprint, aps, draft]{revtex4}
\usepackage{graphicx}

\begin{document}
\title{Cosmology with quantum matter and a classical gravitational field: the approach of configuration-space ensembles}

\author{Marcel Reginatto}
\affiliation{Physikalisch-Technische Bundesanstalt, Bundesallee 100,
38116 Braunschweig, Germany}

\begin{abstract}
I consider the formulation of hybrid cosmological models that consists of a classical gravitational field interacting with a quantized massive scalar field in the formalism of ensembles on configuration space. This is a viable approach that provides an alternative to semiclassical gravity. I discuss a particular, highly nonclassical solution in two approximations, minisuperspace and spherically-symmetric midisuperspace. In both cases, the coupling of the quantum scalar field and classical gravitational field leads to a cosmological model which has a quantized radius of the universe.
\end{abstract}

\maketitle

\section{Introduction}

There are a few good reasons to consider hybrid systems in which the gravitational field remains classical while matter is assumed to consist of quantum fields. A full theory of quantum gravity is not yet available, and an approximation in which spacetime remains classical while matter is described in terms of quantum fields is often physically and computationally appropriate. Furthermore, since the quantization of gravity does not appear to follow from consistency arguments alone \cite{AKR08}, it is of interest to investigate to what extent a hybrid system may provide a consistent, satisfactory description of matter and gravitation. The study of such systems can provide valuable clues that may help in the search for a full quantum theory of gravity. Finally, one must also consider the possibility that the gravitational field may not be quantum in nature \cite{R63,C08,B09}. For example, Butterfield and Isham, while putting forward the point of view that some type of theory of quantum gravity should be sought, have concluded that there is arguably no definitive proof that general relativity has to be quantized \cite{BI01}. Dyson has argued that it might be impossible in principle to observe the existence of individual gravitons, and this has lead him to the conjecture that ``the gravitational field described by Einstein's theory of general relativity is a purely classical field without any quantum behaviour'' \cite{D04}. His observations regarding the impossibility of detecting gravitons have been confirmed by detailed calculations \cite{RB06,BR06}. If Dyson's conjecture is true, hybrid models become unavoidable.

In the standard approach used for coupling quantum fields to a classical gravitational field (i.e., semiclassical gravity), the energy momentum tensor that serves as the source in the Einstein equations is replaced by the expectation value of the energy momentum operator $\widehat{T}_{\mu \nu }$ with respect to some quantum state $\Psi $:
\begin{equation}
^{4}R_{\mu \nu }-\frac{1}{2}g_{\mu \nu }\,^{4}R+\lambda g_{\mu \nu }=\frac{
\kappa }{2}\,\left\langle \Psi \right\vert \widehat{T}_{\mu \nu }\left\vert
\Psi \right\rangle \nonumber
\end{equation}
where $^{4}R_{\mu \nu }$ is the curvature tensor, $^{4}R$\ the curvature scalar and $g_{\mu \nu }$ the metric tensor in spacetime, $\lambda $ is the cosmological constant and $\kappa =16\pi G$ (in units where $c=1$ and $G$ is the gravitational constant). This approach, however, presents a number of well known difficulties. An alternative to semiclassical gravity is provided by the formalism of ensembles on configuration space \cite{HR05}. The formalism can be applied to both quantum and classical systems and allows a general and consistent description of interactions between them. When applied to hybrid systems, it can be shown that the approach overcomes difficulties arising in previous attempts; in particular, the correct equations of motion for the quantum and classical sectors are recovered in the limit of no interaction, conservation of probability and energy are satisfied, uncertainty relations hold for conjugate quantum variables, and the formalism allows a back reaction of the quantum system on the classical system \cite{H08,RH09}. The approach can be generalized to describe the coupling of a quantized field to a classical gravitational field; it is therefore an appropriate formalism to use if one wants to investigate hybrid cosmological models.

In the next section, I describe the coupling of a quantized scalar field to a classical gravitational field using ensembles on configuration space. I then consider a particular, highly nonclassical solution in two approximations, minisuperspace and spherically-symmetric midisuperspace. In the last section, I summarize the results and give conclusions. Technical details are provided in the three Appendices.

\section{Equations for a quantized scalar field interacting with a classical metric field}

In the approach followed here, the interaction of quantum matter with a classical gravitational field will be described using the formalism of ensembles on configuration space. A description of physical systems by means of ensembles on configuration space may be introduced at a very fundamental level, requiring only the notions of probability and an action principle. Consider the situation in which the configuration of a physical system is subject to uncertainty. Under these circumstances, the system must be described by an ensemble of configurations with probability $P$. Introduce an action principle to specify the dynamics. In the Hamiltonian formulation, this implies the existence a function, call it $S$, that is conjugate to $P$, a symplectic structure with its corresponding Poisson bracket algebra, and an ensemble Hamiltonian that is the generator of time evolution. The state of a given system is therefore completely described by the pair of functions $P$ and $S$ and time evolution is determined by the choice of ensemble Hamiltonian.

Since the formalism of ensembles on configuration space is not widely known, a detailed, self-contained introduction is given in \ref{AppECS}, where the approach is illustrated using ensembles that describe classical, quantum and hybrid classical-quantum systems of particles \cite{HR05,H08,RH09}. In this section, the focus will be on the application of the formalism to the coupling of a quantized scalar field to a classical gravitational field. I first illustrate the approach for the case of vacuum gravity \cite{HKR03,H05,R05} and then discuss the formulation for hybrid systems.

The most direct way of introducing a classical configuration space ensemble for gravity is to start from the Einstein-Hamilton-Jacobi equation, which in the metric representation takes the form \cite{MTW73}
\begin{equation}\label{ehj}
\mathcal{H}_{h}=\kappa G_{ijkl}\frac{\delta S}{\delta h_{ij}}\frac{\delta S}{
\delta h_{kl}}-\frac{1}{\kappa }\sqrt{h}\left( R-2\lambda \right) =0,
\end{equation}
where $R$\ is the curvature scalar and $h_{kl}$ the metric tensor on a three-dimensional spatial hypersurface, and $G_{ijkl}=\left( 2h\right) ^{-1/2}\left( h_{ik}h_{jl}+h_{il}h_{jk}-h_{ij}h_{kl}\right) $ is the DeWitt supermetric. The functional $S$ is assumed to be invariant under the gauge group of spatial coordinate transformations, which is equivalent to satisfying the momentum constraints of the canonical formulation of general relativity. Eq.~(\ref{ehj}) corresponds to an infinity of constraints, one at each point. It is possible to introduce an alternative viewpoint \cite{G93, K93} in which Eq.~(\ref{ehj}) is regarded as an equation to be integrated with respect to a ``test function'' in which case we are dealing with one equation for each choice of lapse function $N$,
\begin{equation} \label{hg}
\int d^{3}x\,\,N \mathcal{H}_{h}=0;
\end{equation}
i.e. for each choice of foliation. Such an alternative viewpoint is extremely useful: although it may be impossible to find the general solution (which requires solving the Einstein-Hamilton-Jacobi equation for all choices of lapse functions), it may be possible to find particular solutions for specific choices; for example, the choice $S \sim \int d^{3}x\, \sqrt{h}$ is a particular solution that describes de Sitter spacetime in a flat foliation \cite{K93}.

An appropriate ensemble Hamiltonian for vacuum gravity is given by
\begin{equation} \label{eehj}
\tilde{H}_{h}=\int d^{3}x\int Dh\,P\,\,N \mathcal{H}_{h}
\end{equation}
(technical issues are discussed in more detail in \ref{CEGF} and in ref. \cite{R05}). The functional $P$ is also assumed, like $S$, to be invariant under the gauge group of spatial coordinate transformations. The corresponding equations have the form
\begin{equation}
\frac{\partial P}{\partial t}=\frac{\delta \tilde{H}_{h}}{\delta S},\quad
\frac{\partial S}{\partial t}=-\frac{\delta \tilde{H}_{h}}{\delta P}
\end{equation}
where $\delta /\delta F$ denotes the variational derivative with respect to a functional $F$ \cite{HKR03}. Assuming the constraints $\frac{\partial S}{\partial t}=$ $\frac{\partial P}{\partial t}=0$, these equations lead to Eq.~(\ref{hg}), as required, and to a continuity equation,
\begin{equation} \label{cehj}
\int d^{3}x \,N \frac{\delta }{\delta h_{ij}}\left( P\,G_{ijkl}\frac{\delta S}{
\delta h_{kl}}\right) =0.
\end{equation}
These two equations define an ensemble on configuration space for the case of vacuum gravity.

A hybrid system where a quantum scalar field $\phi$ couples to the classical metric $h_{kl}$ requires a generalization of Eq.~(\ref{eehj}) in which \cite{HR05}
\begin{equation} \label{genphi}
\tilde{H}_{\phi h}=\int d^{3}x\int Dh\,P \,N \left[ \mathcal{H}_{\phi h}+F_{\phi }
\right] ,
\end{equation}
where
\begin{equation}
\mathcal{H}_{\phi h}=\mathcal{H}_{h}+\frac{1}{2\sqrt{h}}\left( \frac{\delta S
}{\delta \phi }\right) ^{2}+\sqrt{h}\left[ \frac{1}{2}h^{ij}\frac{\partial
\phi }{\partial x^{i}}\frac{\partial \phi }{\partial x^{j}}+V\left( \phi
\right) \right]
\end{equation}
is the ensemble Hamiltonian for gravity with a classical scalar field and
\begin{equation}
F_{\phi }=\frac{\hbar ^{2}}{4}\frac{1}{2\sqrt{h}}\left( \frac{\delta \log P
}{\delta \phi }\right) ^{2} \nonumber
\end{equation}
is an additional, non-classical kinetic energy term. Assuming again the constraints $\frac{\partial S}{\partial t}=$ $\frac{\partial P}{\partial t}=0$, the coresponding equations are given by
\begin{equation} \label{hh}
\int d^{3}x\, N \, \left[ \mathcal{H}_{\phi h}
-\frac{\hbar ^{2}}{2\sqrt{h}}
\left( \frac{1}{A}\frac{\delta ^{2}A}{\delta \phi ^{2}}\right) \right]=0, \nonumber
\end{equation}
where $A\equiv \sqrt{P}$, and a continuity equation that is identical to Eq.~(\ref{cehj}).

\section{Hybrid cosmological model in minisuperspace}

Before introducing the cosmological model in spherical gravity that is discussed in the next section, it will be instructive to look at the corresponding minisuperspace model \cite{HR05}.

Consider a closed Robertson-Walker universe with a massive scalar field. The line element can be written in the form
\begin{equation}
ds^{2}=-N^{2}\left( t\right) dt^{2}+a^{2}\left( t\right) d\Omega _{3}^{2}
\end{equation}
where $a$ is the scale factor and $d\Omega _{3}^{2}$ is the standard line
element on $S^{3}$. Symmetry reduction leads to a minisuperspace formulation in a finite dimensional configuration space with coordinates $a$ and $\phi$ \cite{K04}. Following Ref. \cite{HR05}, I restrict to a potential term that is quadratic in $\phi $. Then, the classical Hamilton-Jacobi equation takes the form $\mathcal{H}_{\phi a}=0$, with
\begin{equation} \label{CHJE-MSM}
\mathcal{H}_{\phi a}=-\frac{1}{a}\left( \frac{\partial S}{\partial a}\right)
^{2}+\frac{1}{a^{3}}\left( \frac{\partial S}{\partial \phi }\right) ^{2}-a+
\frac{\lambda a^{3}}{3}+m^{2}a^{3}\phi ^{2} ,
\end{equation}
where $m$ is the mass of the field and the units of this section have been chosen so that $2G/3\pi =1$ (to agree with those of Ref. \cite{HR05}).

The corresponding ensemble Hamiltonian for a quantized field interacting with
the classical metric is given by \cite{HR05}
\begin{equation}
\tilde{H}_{\phi a}=\int dad\phi P\left[ \mathcal{H}_{\phi a} + \frac{\hbar^{2}}{4}\frac{1}{a^{3}}\left( \frac{\partial \log P}{\partial \phi }\right) ^{2}\right] .
\end{equation}
Assuming the constraints $\frac{\partial S}{\partial t}=$ $\frac{\partial P}{\partial t}=0$, the equations of motion take the form
\begin{equation}
\mathcal{H}_{\phi a}-\frac{\hbar ^{2}}{a^{3}}\frac{1}{\sqrt{P}}\frac{
\partial ^{2}\sqrt{P}}{\partial \phi ^{2}}=0  \label{HQ}
\end{equation}
and
\begin{equation}
-\frac{\partial }{\partial a}\left( \frac{P}{a}\frac{\partial S}{\partial a}
\right) +\frac{\partial }{\partial \phi }\left( \frac{P}{a^{3}}\frac{
\partial S}{\partial \phi }\right) =0.  \label{C}
\end{equation}

An \textit{exact solution} can be derived for the case $S=0$. This is a highly non-classical solution: Eq. (\ref{CHJE-MSM}), the classical Hamilton-Jacobi equation,  does not admit any solutions with this ansatz. When $S=0$, Eq.~(\ref{HQ}) reduces to
\begin{equation}
-\frac{\hbar ^{2}}{a^{3}}\frac{1}{\sqrt{P}}\frac{\partial ^{2}\sqrt{P}}{
\partial \phi ^{2}}-a+\frac{\lambda a^{3}}{3}+a^{3}m^{2}\phi ^{2}=0
\label{HHJ}
\end{equation}
while Eq.~(\ref{C}) is automatically satisfied. The non-negative,
normalizable solutions take the form
\begin{equation}
P_{n}\left( \phi ,a\right) = \delta \left( a-a_{n}\right) \frac{\alpha_{n}}{\sqrt{\pi }2^{n}n!}\exp \left( -\alpha_{n}^{2}\phi ^{2}\right) \,
\left[H_{n}\left( \alpha_{n}\phi \right)\right]^2   \nonumber
\end{equation}
where the $H_{n}$ are Hermite polynomials, $\alpha_{n}^{2}=a_{n}^{3}m/\hbar \ $and the $a_{n}$\ satisfy the condition
\begin{equation} \label{quanta}
a_{n}-\frac{\lambda a_{n}^{3}}{3}=2\hbar m\left( n+\frac{1}{2}\right)
\end{equation}
for $n=\left\{ 0,1,2,...\right\} $. If the term proportional to the cosmological constant $\lambda $ can be neglected, the quantization condition takes the simple form $a_{n}=2\hbar m\left( n+\frac{1}{2}\right) $.

While the transformation $\psi =\sqrt{P}$ leads, via Eq. (\ref{HHJ}), to a Schr\"{o}dinger equation for $\psi $, it is not possible to introduce solutions that are linear superpositions of the $\psi_{n}$ because the potential term in the equation ends up being a function of $a_{n}$.

This solution that has been derived in this section has some remarkable features. One can see that the coupling of the quantum field to a purely classical metric leads to a {\it quantization condition} for the scale factor $a$. Furthermore, the classical singularity at $a=0$ is \textit{excluded} from these solutions. Finally, notice that there is a natural ordering of the solutions $\left\{P_{n}\right\}$ in terms of $n$ and one may argue that this ordering leads to a thermodynamic arrow of time. This follows from the observation that the amount of {\ structure} associated with a solution $P_{n}$ (as determined, for example, by counting the number of nodes in $\psi_{n}$ or by evaluating the entropy expression $-\int d\phi\, P_{n}\log P_{n}$ for different values of $n$)\ increases with increasing $n$. Note that this
thermodynamic arrow of time coincides with the arrow of time as determined by an expanding universe whenever the non-linear term proportional to $\lambda$ can be neglected.

\section{Hybrid cosmological model in spherically symmetric gravity}

Consider now a midisuperspace hybrid cosmological model in spherically symmetric gravity. In the case of spherical symmetry, the line element may be written in the form
\begin{equation} \label{gssg}
g_{\mu \nu }dx^{\mu }dx^{\nu }=-N^{2}dt^{2}+\Lambda ^{2}\left(
dr+N^{r}dt\right) ^{2}+R^{2}d\Omega ^{2}.
\end{equation}
The lapse function $N$ and the shift function $N^r$ are functions of the radial coordinate $r$ and the time coordinate $t$. The configuration space for the gravitational field consists of two fields, $R$ and $\Lambda$. Under transformations of $r$, $R$ behaves as a scalar and $\Lambda$ as a scalar density. Spherically symmetric gravity is discussed in detail in a number of papers, mostly in reference to the canonical quantization of black hole spacetimes. For discussions using the metric representation, see for example \cite{R95, K94, L95}. For discussions of the Einstein-Hamilton-Jacobi equation in the context of the WKB approximation of quantized spherically symmetric gravity, see for example \cite{FMP90, BK97}.

I now set $\hbar=c=G=1$. The Einstein-Hamilton-Jacobi equation for the case of a vacuum gravity takes the form
\begin{equation}
\mathcal{H}_{\Lambda R} = -\frac{1}{R}\frac{\delta S}{\delta R}\frac{\delta S}{\delta
\Lambda }+\frac{1 \Lambda }{2R^{2}}\left( \frac{\delta S}{\delta \Lambda }
\right) ^{2}+ \lambda\frac{\Lambda R^{2}}{2}+ V=0
\end{equation}
where $S$ is assumed to be invariant under diffeomorphisms. $V$ is related to the curvature scalar by $^{4}R=-4 \, \Lambda R^2 \, V$, and is given by
\begin{equation} \label{v}
V=\frac{RR^{\prime \prime }}{\Lambda }-\frac{RR^{\prime }\Lambda ^{\prime }}{\Lambda ^{2}}
+\frac{R^{\prime 2}}{2\Lambda }-\frac{\Lambda }{2}
\end{equation}
where primes indicate derivatives with respect to $r$.

The ensemble Hamiltonian of a hybrid system where the matter is in the form of a minimally coupled quantized radially symmetric scalar field of mass $m$ is given by
\begin{equation} \label{HphiLambdaR}
\tilde{H}_{\phi \Lambda R}=\int dr\int Dh\,P\,N \left[ \mathcal{H}_{\phi \Lambda R}+F_{\phi }
\right] ,
\end{equation}
where
\begin{equation}
\mathcal{H}_{\phi \Lambda R}=\mathcal{H}_{\Lambda R}
+\frac{1}{2\Lambda R^{2}}\left( \frac{\delta S}{\delta \phi }\right) ^{2}
+\frac{R^2 }{2 \Lambda}\phi ^{\prime 2} + \frac{\Lambda R^{2} m^2}{2 } \phi ^{ 2} ,
\end{equation}
and
\begin{equation}
F_{\phi }= \frac{1}{8 \Lambda R^2}\left( \frac{\delta \log P
}{\delta \phi }\right) ^{2}.
\end{equation}
Eq.~(\ref{HphiLambdaR}) is the analogous of Eq.~(\ref{genphi}) for the case of spherically symmetric gravity. Assuming again the constraints $\frac{\partial S}{\partial t}=$ $\frac{\partial P}{\partial t}=0$, the corresponding equations are
\begin{equation} \label{ehjssg}
\int dr\, N \, \left[ \mathcal{H}_{\phi \Lambda R}
-\frac{1}{2\Lambda R^2}
\left( \frac{1}{A}\frac{\delta ^{2}A}{\delta \phi ^{2}}\right) \right]=0,
\end{equation}
where $A\equiv P^{1/2}$, and the continuity equation
\begin{equation} \label{cssg}
\int dr\, N \, \left[
\frac{\delta }{\delta R}\left( P\frac{1}{R}\frac{\delta S}{\delta \Lambda }\right)
+\frac{\delta }{\delta \Lambda }\left( P\frac{1}{R}\frac{\delta S}{\delta R}
-P\frac{\Lambda }{R^{2}}\frac{\delta S}{\delta\Lambda }\right)
-\frac{\delta }{\delta \phi }\left( P\frac{1}{\Lambda R^{2}}
\frac{\delta S}{\delta \phi }\right) \right]=0.
\end{equation}

I now want to consider a class of solutions that is analogous to the class of minisuperspace solutions described in the previous section. To do this, I will look for a solution that satisfies the following requirements:
\begin{enumerate}
\item it corresponds to choosing a foliation of spaces of constant positive curvature and a lapse function $N$ that is constant, and
\item it satisfies $S=0$.
\end{enumerate}
In this case, Eqs. (\ref{ehjssg}) and (\ref{cssg}) reduce to a single Schr\"{o}dinger functional equation for $A$; i.e., of the type
\begin{equation}
-\frac{1}{2\Lambda R^2}
\left(\frac{\delta ^{2}A}{\delta \phi ^{2}}\right)+ \left[ \lambda\frac{\Lambda R^2}{2} + V +\frac{R^2 }{2 \Lambda}\phi ^{\prime 2} + \frac{\Lambda R^{2} m^2}{2 } \phi ^{ 2} \right] A = 0.
\end{equation}
This equation can be solved using standard techniques developed for the
Schr\"{o}dinger functional representation of quantum field theory \cite{K04,J90,H92,LS98}.

To get an explicit solution that corresponds to the lowest state of the minisuperspace model that I considered in the previous section, I will assume that $A$ has the form of a ground state Gaussian functional; i.e.,
\begin{equation}
A_{(0)} \sim \exp\left\{ -\frac{1}{2} \int \int dy \, dz \, \Lambda_y \, \Lambda_z \, R_y^2 \, R_z^2 \; \phi_y \, K_{yz} \, \phi_z \right\} \nonumber.
\end{equation}
Instead of $A_{(0)}$, one may also consider the excited states which solve the functional Schr\"{o}dinger equation; I discuss the consequences of making this alternative choice at the end of this section.

As shown in \ref{CSGFS}, the equation that determines the functional $A_{(0)}$ can be mapped to a functional Schr\"{o}dinger equation in a space of constant curvature and the kernel $K_{xy}$ can be expressed in the simple form
\begin{equation}\label{Kxy}
K_{xy} = \frac{1}{2 a_0^4} \sum_n \sqrt{\gamma_n } \, \psi^{(n)}_x \psi^{(n)}_y,
\end{equation}
where the $\psi^{(n)}_r$ are solutions of a Schr\"{o}dinger-type equation in a space of constant curvature,
\begin{equation}
-\frac{1}{\sin^2r}\,\frac{\partial}{\partial r} \left( \sin^2 r \, \frac{\partial \psi_r^{(n)}}{\partial r} \right) +  m^2 a_0^2 \, \psi_r^{(n)} = \gamma_n \psi_r^{(n)}.
\end{equation}
The eigenvalues $\gamma_n$ are given by
\begin{equation}
\gamma_n = n^2 - 1 + m^2 a_0^2,~~~~~n=1,2,3...~~.
\end{equation}

As described in \ref{CSGFS}, in principle it is possible to get an expression for $a_0$ that depends on the cosmological constant $\lambda$ and the energy $E_{(0)}$ of the quantized scalar field, which is given by \cite{LS98}
\begin{equation}
E_{(0)} \sim a_0^3 \int dr\, \sin^2r \, K_{rr}.
\end{equation}
However, $\int dr \, \sin^2r \, K_{rr} \sim \sum_{n} \gamma_n$, which diverges. This is a consequence of the infinite zero-point energy of the quantum field. It is necessary to use a renormalization procedure to extract a finite result.

In this particular case, the equation for $A_{(0)}$ is similar in form to a functional Schr\"{o}dinger equation for a quantum scalar field in an Einstein universe, so it is possible to use previous results from the literature where such renormalization procedures have been carried out (for a thorough analysis, see reference \cite{HRS08}). Note that this is not the generic case: since the equations of hybrid cosmology are not linear, in general they will not map to a Schr\"{o}dinger-type functional equation. The simplification that is achieved here is a direct consequence of choosing $S=0$. With a different ansatz, the equations can not be solved in this way. A discussion of renormalization procedures for this solution is outside of the scope of this paper and will be the subject of a future publication.

A similar analysis may be carried out where the ground state functional $A_{(0)}$ is replaced by an {\it excited} state. Consider a first excited state $A_{(1)}$ specified by the eigenfunction $\psi_r^{(n)}$. This state will differ in energy from the (divergent) ground state energy by a finite amount $\Delta E(n)$ which depends on the value of $\gamma_n$, with $\Delta E(n) = \sqrt{\gamma_n}$ \cite{LS98}. $\Delta E(n)$ can only take discrete values because $\gamma_n$ is quantized, and this means that $a_0$ will also be restricted to {\it discrete values}. Therefore, the coupling of the quantum scalar field and classical gravitational field leads to the quantization of the radius of the universe, not only for the minisuperspace model but also for the midisuperspace model.

\section{Concluding remarks}

The main result of this paper is a formulation of hybrid cosmological models that makes use of the formalism of ensembles on configuration space and thus provides an alternative to other approaches (e.g., semiclassical gravity) which couple quantum matter to classical gravitational fields. In particular, I have examined the case in which a classical gravitational field interacts with a quantized massive scalar field. I have derived a particular, highly nonclassical solution using two approximations, minisuperspace and spherically-symmetric midisuperspace.

The hybrid cosmological model examined here was previously considered in reference \cite{HR05} but only within the minisuperspace approximation. One of the aims of the present work was to reexamine this model using the next level of approximation; i.e., in the context of a midisuperspace formulation like spherically symmetric gravity. It appears that the space of solutions in midisuperspace may be very rich by comparison. This is a consequence of the renormalization procedure that needs to be carried out to achieve a finite result. This complexity is absent from the minisuperspace solution.

In both minisuperspace and midisuperspace, the coupling of the quantum scalar field to the classical gravitational field leads to the quantization of the radius of the universe, a remarkable result. It would be interesting to know whether this is a generic feature of hybrid closed cosmological models or whether this is a feature that is particular to the family of solutions considered here (i.e., those with $S=0$). In the minisuperspace solution, it is clear that a singular solution with $a=0$ is excluded; the situation is more complicated in the case of the midisuperspace solution due to the need to introduce renormalization.

In the Einstein universe of classical relativity, the cosmological constant acts as a repulsive force which balances the gravitational attraction of matter; it is essential to include it in order to achieve a static solution. In the hybrid solutions considered here, it is possible to achieve static solutions with a vanishing cosmological constant and it would appear that quantum fluctuations are sufficient to balance the gravitational attraction. It would be interesting to extend these results to consider perturbative solutions without the restriction $S=0$, which should describe solutions that are not static.

\section{Acknowledgments}

I thank M. J. W. Hall and Stephen Boughn for valuable discussions.

\appendix
\section{Ensembles on configuration space}\label{AppECS}

This appendix provides a brief summary of the formalism of ensembles on configuration space. Derivations of the results stated in this appendix can be found in Refs. \cite{HR05, RH09, H08}.

\subsection{Ensembles on configuration space describing classical,
quantum and mixed classical-quantum systems}

Start with the assumption that, as seems to be implied by quantum mechanics, the configuration of a physical
system is an inherently statistical concept. The system is therefore
described by an ensemble of configurations with probability density
$P$, where $P \geq 0$ and $\int dx \,P(x,t) = 1 $. To derive
equations of motion, introduce an {\it ensemble Hamiltonian}
$\tilde{H}[P,S]$, where $S$ is an auxiliary field that is
canonically conjugate to $P$. The equations of motion take the form
\begin{equation}
\frac{\partial P}{\partial t} = \left\{ P,\tilde{H} \right\}_{PB}
= \frac{\delta\tilde{H}}{\delta S},~~~~~~~~~\frac{\partial S}{\partial t} = \left\{ S,\tilde{H} \right\}_{PB}
=-\frac{\delta\tilde{H}}{\delta P},
\end{equation}
where $\{ A,B \}_{PB}$ is the Poisson bracket of the fields $A$ and $B$.

The following ensemble Hamiltonians lead to equations that describe
the evolution of quantum and classical non-relativistic systems,
\begin{eqnarray}\label{HCandQ_App}
\tilde{H}_C[P,S] &=& \int dx\, P \left[ \frac{|\nabla S|^2}{2m} + V(x)\right] ,\nonumber\\
\tilde{H}_Q[P,S] &=& \tilde{H}_C[P,S]
+  \frac{\hbar^2}{4} \int dx\ P\frac{|\nabla \log P|^2}{2m} .
\end{eqnarray}
For example, the equations of motion derived from $\tilde{H}_Q[P,S]$ are given by
\begin{equation}\label{QEqMotion_App}
\frac{\partial P}{\partial t} + \nabla .\left( P\frac{\nabla S}{m} \right) =0,~~~\frac{\partial S}{\partial t} + \frac{|\nabla S|^2}{2m} + V +  \frac{\hbar^2}{2m}\frac{\nabla^2 P^{1/2}}{P^{1/2}} = 0
\end{equation}
while the equations of motion derived from $\tilde{H}_C[P,S]$ are
the same as Eq. (\ref{QEqMotion_App}) but with $\hbar=0$. The first
equation in Eq. (\ref{QEqMotion_App}) is a continuity equation, the
second equation is the classical Hamilton-Jacobi equation when
$\hbar = 0$ and a modified Hamilton-Jacobi equation when $\hbar \neq
0$. Defining $\psi=\sqrt{P}~e^{iS/\hbar}$, Eq.
(\ref{QEqMotion_App}) takes the form
\begin{equation}\nonumber
i\hbar \frac{\partial \psi}{\partial t}
= \frac{-\hbar^2}{2m}\nabla^2\psi + V\psi,
\end{equation}
which is the usual form of the Schr\"{o}dinger equation. Therefore,
in this formalism, quantum and classical particles are treated on an
equal footing, with differences being primarily due to the different
forms of the respective ensemble Hamiltonians.

It is possible to extend the formalism in a natural way to allow for
mixed quantum-classical systems. A mixed quantum-classical ensemble
Hamiltonian on a configuration space with coordinates $q$, $y$ is
given by
\begin{eqnarray}\label{HQC_App}
\tilde{H}_{QC}[P,S] &=& \int dq\,dy\, P\,\left[ \frac{|\nabla_y S|^2}{2M}
+ \frac{|\nabla_q S|^2}{2m}
 \right] \nonumber\\
&+& \int dq\,dy\, P\,\left[  \frac{\hbar^2}{4} \frac{|\nabla_q \log P|^2}{2m} + V(q,y,t)\right].
\end{eqnarray}
Here $q$ denotes the configuration space coordinate of a quantum
particle of mass $m$ and $y$ that of a classical particle of mass
$M$, and $V(q,y,t)$ is a potential energy function describing the
quantum-classical interaction. The equations of motion for $P$ and
$S$ derived from $\tilde{H}_{QC}$ are
\begin{eqnarray}\label{EqsCQ_App}
\frac{\partial P}{\partial t} &=&  -\nabla_q .\left( P
\frac{\nabla_q S}{m} \right) - \nabla_y .\left(P\frac{\nabla_y S}{M}\right), \nonumber\\
\frac{\partial S}{\partial t} &=& -
\frac{|\nabla_q S|^2}{2m} - \frac{|\nabla_y S|^2}{2M} - V +
\frac{\hbar^2}{2m}\frac{\nabla_q^2 P^{1/2}}{P^{1/2}} .
\end{eqnarray}

The state of a system is described by specifying the two fields $P$
and $S$. While the interpretation of $P$ is straightforward, there
are some subtle issues concerning the physical interpretation of
$S$. For the cases discussed here (i.e., classical,
quantum and mixed systems), it is possible to define local energy
and momentum densities in terms of $S$. If $\tilde{H}[\lambda P, S]
= \lambda \tilde{H}[P,S]$ (which holds for the ensemble Hamiltonians
that we considered above), then
\begin{equation}
\tilde{H} = \int dx \,P\frac{\delta\tilde{H}}{\delta P} = - \int
dx\, P\frac{\partial S}{\partial t} = - \langle \partial S/\partial
t \rangle ,
\end{equation}
which shows that $-P \partial S/\partial t$ is a local energy density.
Furthermore, $\int dx\, P\nabla S$ is the canonical infinitesimal
generator of translations, since
\begin{eqnarray}
\delta P(x) = \delta \textbf{x} \cdot \left \{ P, \int dx\,
P\nabla S \right \}_{PB} = - \delta \textbf{x} \cdot \nabla P ,
\nonumber\\
\delta S(x) = \delta \textbf{x} \cdot \left \{ S, \int dx\,
P\nabla S \right \}_{PB} = - \delta \textbf{x} \cdot \nabla S ,
\end{eqnarray}
under action of the generator, and therefore $P\nabla S$ can be
considered a local momentum density. To maintain full generality, $S$
should not be regarded as a ``momentum potential''. In particular,
for an ensemble of classical particles with uncertainty described by
the probability $P$, it will not be assumed that the momentum of a
member of the ensemble is a well-defined quantity proportional to
the gradient of $S$, as it is done in the usual deterministic
interpretation of the Hamilton--Jacobi equation. This avoids forcing
a similar deterministic interpretation in the quantum and
quantum-classical cases. A deterministic picture can be recovered
for classical ensembles precisely in those cases in which
trajectories are operationally defined.

\subsection{Observables and local densities}

Observables are functionals of $P$ and $S$. Given two functionals
$A[P,S]$ and $B[P,S]$, define their Poisson bracket in the
standard way,
\begin{equation}
\{ A,B \}_{PB} = \int dx \left( \frac{\delta A}{\delta P} \frac{\delta B}{\delta S} - \frac{\delta A}{\delta S} \frac{\delta B}{\delta P} \right) ,
\end{equation}
which gives an algebra of obervables.

Arbitrary functionals $A[P,S]$ are not necessarily observables
because observables have to satisfy certain mild requirements. For
example, the infinitesimal canonical transformation generated by any
observable $A$ must preserve the normalization and positivity of
$P$. This implies the two conditions
\begin{equation}
A[P,S+c] = A[P,S],~~~~~\delta A / \delta S = 0 ~ \textrm{if} ~
P(x)=0.
\end{equation}
Note that the first equation implies gauge invariance of the theory under $S
\rightarrow S + c$.

A more general condition that may be imposed on observables is that
they be homogenous of degree one in $P$; i.e., $A[\lambda
P,S]=\lambda A[P,S]$. This is a condition that is satisfied by both
classical and quantum systems, and it will be assumed here that it is also
valid for mixed classical-quantum systems. Then, it follows that
\begin{equation}
A[P,S] = \int dx \,P \left( \delta A / \delta P \right) = \langle \delta A/ \delta P \rangle .
\end{equation}
That is, one can associate with each observable $A$ a local density
on the configuration space, and the value of $A$ can be calculated
by integrating over this local density.

As a simple example, consider position and momentum observables.
In all three cases (classical, quantum, and mixed
classical-quantum), they are given by $X[P,S]=\int dx Px$ and
$\Pi[P,S]= \int dx P \nabla S$, with corresponding local densities
$Px$ and $P \nabla S$. Such an equivalence does not hold of course
for more complicated observables. For example, different functionals
are needed to describe the kinetic energy of classical, quantum, and
mixed classical-quantum systems, as is apparent from their
corresponding ensemble Hamiltonians.

\subsection{An example of a mixed classical/quantum system: Measurement of the position of a quantum particle carried out by a classical apparatus}\label{CQmeas}

To illustrate the application of the formalism, I consider a simple model: a classical apparatus which measures the position of a quantum particle \cite{HR05}.

To model a measurement of position, introduce the ensemble
Hamiltonian
\begin{equation}
\tilde{H}_{\rm position} = \tilde{H}_{QC} + \kappa(t) \int dq\,dx\, P\, q.\nabla_x S .
\end{equation}
For an interaction over a short time period $[0,T]$ such that
$\tilde{H}_{QC}$ can be ignored during the interaction,
\begin{equation}
\frac{\partial P}{\partial t} = -\kappa(t)\, q.\nabla_x P ,
~~~~~~ \frac{\partial S}{\partial t} = -\kappa(t)\, q.\nabla_x S ,
\end{equation}
which integrates to ($K=\int_0^T dt\,\kappa(t)$)
\begin{equation}
P(q,x,T) = P(q, x-Kq, 0),~~~~~~S(q,x,T) = S(q, x-Kq,0) .
\end{equation}
Consider the case where the initial position of the pointer is
sharply defined, $P(q,x,0) = \delta(x-x_0)P_Q(q)$. Then, after the
measurement, $\Delta x \neq 0$, with probability $P_Q(q)$ of finding
$x=x_0-Kq$. In other words, \emph{uncertainty is transferred} from
the quantum ensemble to the classical ensemble. Transfer of
uncertainty is the norm whenever two systems interact.

To interpret the outcome it will be helpful to introduce some
additional concepts. Consider the conditional
probability $P(q|x) = P(q,x)/P(x) $. The conditional wave function
is defined by
\begin{equation}
\psi(q|x) = \sqrt{P(q|x)}\, e^{i S(q,x)/\hbar}
,~~~~~~~|\psi_x\rangle = \int dq\,\psi(q|x)\,|q\rangle .
\end{equation}
Introduce a conditional density operator,
\begin{equation}
\rho_{Q|C} = \int dx\, P(x)\,|\psi_x\rangle\langle \psi_x| .
\end{equation}
It is important to keep in mind that $\psi(q|x)$ and $\rho_{Q|C}$ do
\textit{not} satisfy linear Schr\"{o}dinger and Liouville equations,
nor unitary invariance properties. These quantities only contain
\textit{partial} information.

The
conditional density for the quantum component is diagonal in the
position basis,
\begin{equation}
\rho_{Q|C} = \int dq \,P_Q(q)\, |q\rangle\langle q|
\end{equation}
and thus ``decoheres'' with respect to position.

The main features of this simple measurement model are: (i) the measuring apparatus
is described classically; (ii) information about quantum ensembles is
obtained via an appropriate interaction with an ensemble of
classical measuring apparatuses, which correlates the classical
configuration with a corresponding quantum property, and (iii) there
is a conditional decoherence of the quantum ensemble relative to the
classical ensemble, which depends upon the nature of the
quantum-classical interaction.

\subsection{Entanglement}

Entangled states in quantum mechanics are
those states of a multi-particle state that can not be expressed as
a product of single-particle states. A non-entangled state is
therefore one that has a wave function $\psi$ that can be written in
the form
\begin{equation}\label{nonEntQuantState}
\psi(t,x_1,x_2,...,x_n)=\prod_{k=1}^{n}{\psi_k(t,x_k)}
\end{equation}
where the $\psi_k$ are single-particle wave functions. In other
words, a non-entangled state is one in which $P$ and $S$ satisfy the
conditions
\begin{eqnarray}\label{nonEnt}
P(t,x_1,x_2,...,x_n)=\prod_{k=1}^{n}{P_k(t,x_k)},\nonumber \\
S(t,x_1,x_2,...,x_n)=\sum_{k=1}^{n}{S_k(t,x_k)}
\end{eqnarray}

There is no difficulty in extending this distinction between
entangled and non-entagled states to physical states described in
the ECS formalism; in particular, classical and mixed
classical-quantum states can be entangled since they need not
satisfy the requirements of Eq. (\ref{nonEnt}). Therefore,
entanglement in this sense is a general feature of ensembles in configuration
space and it is not restricted to quantum mechanical systems only.

\section{Classical ensembles of gravitational fields}\label{CEGF}

A Hamilton-Jacobi formulation for the gravitational field can be defined in
terms of the functional equations
\begin{eqnarray}
\mathcal{H} &=& \kappa G_{ijkl}\frac{\delta S}{\delta h_{ij}}\frac{\delta S}{\delta h_{kl}}-\frac{1}{\kappa }\sqrt{h}\left( R-2\lambda \right) =0,  \nonumber\\
\mathcal{H}_{i} &=& D_{j}\left( h_{ik}\frac{\delta S}{\delta h_{kj}}\right) =0,
\end{eqnarray}
where $R$\ is the curvature scalar and $h_{kl}$ the metric tensor on a three-dimensional spatial hypersurface, $\lambda $ is the cosmological constant, $\kappa =16\pi G$ (in units where $c=1$ and $G$ is the gravitational constant), and $G_{ijkl}=\left( 2h\right) ^{-1/2}\left( h_{ik}h_{jl}+h_{il}h_{jk}-h_{ij}h_{kl}\right) $ is the DeWitt supermetric \cite{MTW73}.

As a consequence of the Hamiltonian constraint $\mathcal{H}=0$ and the momentum constraints $\mathcal{H}_{i}=0$,
$S$ must satisfy an infinite set of constraints, numbering four per
three-dimensional point.\ $S$ also satisfies the condition $\frac{\partial S
}{\partial t}=0$ \cite{B66}. The momentum constraints are equivalent to
requiring the invariance of the Hamilton-Jacobi functional $S$ under spatial
coordinate transformations.\ One may therefore formulate the theory in an
equivalent way by keeping the Hamiltonian constraint, ignoring the momentum
constraints, and requiring instead that $S$ be invariant under the gauge
group of spatial coordinate transformations.

To define classical ensembles for gravitational fields, it is necessary to
introduce some additional mathematical structure: a measure $Dh$ over the
space of metrics $h_{kl}$ and a probability functional $P\left[ h_{kl}\right]
$. A standard way of defining the measure \cite{DW79,HW99} is to introduce
an invariant norm for metric fluctuations that depends on a parameter $\omega $,
\begin{equation}
\left\Vert \delta h\right\Vert ^{2}=\int d^{n}x\left[ h\left( x\right)
\right] ^{\omega /2}H^{ijkl}\left[ h(x);\omega \right] \delta h_{ij}\delta
h_{kl}
\end{equation}
where $n$ is the number of dimensions and
\begin{equation}
H^{ijkl}=\frac{1}{2}\left[ h\left( x\right) \right] ^{\left( 1-\omega
\right) /2}\left[ h^{ik}h^{jl}+h^{il}h^{jk}+\lambda h^{ij}h^{kl}\right]
\end{equation}
is a generalization of the inverse of the DeWitt supermetric (in \cite{DW79}
the particular case $\omega =0$ was considered).\ This norm induces a local
measure for the functional integration given by
\begin{equation}
\int d\mu \left[ h\right] =\int \prod_{x}\left[ \det H\left( h\left(
x\right) \right) \right] ^{1/2}\prod_{i\geq j}dh_{ij}\left( x\right)
\end{equation}
where
\begin{equation}
\det H\left( h\left( x\right) \right) \propto \left( 1+\frac{1}{2}\lambda
n\right) \left[ h\left( x\right) \right] ^{\sigma },
\end{equation}
and $\sigma =\left( n+1\right) \left[ \left( 1-\omega \right) n-4\right] /4$
\ (one needs to impose the condition $\lambda \neq -2/n$, otherwise the
measure vanishes). Therefore, up to an irrelevant multiplicative constant,
the measure takes the form
\begin{equation}
\int d\mu \left[ h\right] =\int \prod_{x}\left[ \sqrt{h\left( x\right) }
\right] ^{\sigma }\prod_{i\geq j}dh_{ij}\left( x\right) .  \label{Eq07}
\end{equation}
Without loss of generality, one may set $Dh$ equal to $d\mu \left[ h\right] $
with$\ \sigma =0$, since a term of the form $\left[ \sqrt{h\left( x\right) }
\right] ^{\sigma }$ may be absorbed into the definition of $P\left[ h_{kl}
\right] $.

It is natural to require also that $\int DhP$ be invariant under the gauge
group of spatial coordinate transformations. Since the family of measures
defined by eq. (\ref{Eq07}) is invariant under spatial coordinate
transformations \cite{HW99,FP74}, the invariance of $\int DhP$ leads to a
condition on $P$ that is similar to the one required of $S$. To show this,
consider an infinitesimal change of coordinates $x^{\prime k}=x^{k}+\epsilon
^{k}\left( x\right) $ and the corresponding transformation of the metric, $
h_{kl}\rightarrow h_{kl}-\left( D_{k}\epsilon _{l}+D_{l}\epsilon _{k}\right)
$. The variation of $\int DhP$ can be expressed as
\begin{equation}
\delta _{\epsilon }\int DhP=\int Dh\left[ D_{k}\left( \frac{\delta P}{\delta
h_{kl}}\right) \epsilon _{l}+D_{l}\left( \frac{\delta P}{\delta h_{kl}}
\right) \epsilon _{k}\right] .
\end{equation}
Therefore, $\delta _{\epsilon }\int DhP=0$ requires
\begin{equation}
D_{k}\left( \frac{\delta P}{\delta h_{kl}}\right) =0.  \label{Eq08}
\end{equation}
or the gauge invariance of $P$. In addition to eq. (\ref{Eq08}), it will be
assumed that $\frac{\partial P}{\partial t}=0$ also holds.

Finally, it should be pointed out that one must factor out the infinite
diffeomorphism gauge group volume out of the measure to calculate finite
averages using the measure $Dh$ and probability functional $P$. This can be
achieved using the geometric approach described in \cite{M95}. This issue
will not be discussed further here, since it does not affect the derivation
of the equations of motion.

An appropriate ensemble Hamiltonian for the gravitational field is given by
\begin{equation}
\tilde{H}_{c}=\sum_{x}\int DhP\mathcal{H}\sim \int d^{3}x\int DhP\mathcal{H}.
\label{Eq09}
\end{equation}
The equations of motion derived from eq. (\ref{Eq09})\ are of the form
\begin{equation}
\frac{\partial P}{\partial t}=\frac{\Delta \tilde{H}_{c}}{\Delta S},\quad
\frac{\partial S}{\partial t}=-\frac{\Delta \tilde{H}_{c}}{\Delta P}
\end{equation}
where $\Delta /\Delta F$ denotes the variational derivative with respect to
the functional $F$. With $\frac{\partial S}{\partial t}=$ $\frac{\partial P}{
\partial t}=0$, the equations of motion take the form
\begin{equation}
\mathcal{H}=0,  \label{Eq10}
\end{equation}
and
\begin{equation}
\int d^{3}x\frac{\delta }{\delta h_{ij}}\left( PG_{ijkl}\frac{\delta S}{
\delta h_{kl}}\right) =0.  \label{Eq11}
\end{equation}
Eq. (\ref{Eq10}) is the Hamiltonian constraint, and eq. (\ref{Eq11}) may be
interpreted as a continuity equation.

It is of interest that the interpretation of eq. (\ref{Eq11}) as a
continuity equation leads to a rate equation that relates $\frac{\partial
h_{kl}}{\partial t}$ and $\frac{\delta S}{\delta h_{kl}}$. This follows from
the observation that such an interpretation is only possible if the
\textquotedblleft field velocity\textquotedblright\ $\frac{\partial h_{kl}}{
\partial t}$\ is related to $G_{ijkl}\frac{\delta S}{\delta h_{kl}}$ in a
linear fashion. Indeed, the most general rate equation for the metric $
h_{ij} $ that is consistent with the interpretation of eq. (\ref{Eq11}) as a
continuity equation is of the form
\begin{equation}
\delta h_{ij}=\left( \alpha G_{ijkl}\frac{\delta S}{\delta h_{kl}}+\delta
_{\epsilon }h_{ij}\right) \delta t
\end{equation}
where $\alpha $ is an arbitrary function of $x$ (I have include a term $
\delta _{\epsilon }h_{kl}=-\left( D_{k}\epsilon _{l}+D_{l}\epsilon
_{k}\right) $ which allows for gauge transformations of $h_{kl}$, which is
permitted because the gauge transformations are assumed to leave $\int DhP$
invariant, as discussed before). \ Therefore, one may write the rate
equation for $h_{kl}$ in the standard form
\begin{equation}
\frac{\partial h_{ij}}{\partial t}=NG_{ijkl}\frac{\delta S}{\delta h_{kl}}
+D_{i}N_{j}+D_{j}N_{i}.  \label{Eq13}
\end{equation}
Eq. (\ref{Eq13}) agrees with the equations derived from the ADM canonical
formalism, provided $N$ is identified with the lapse function and $N_{k}$
with the shift vector \cite{W84}.

It is remarkable that the rate equations for the metric, eq. (\ref{Eq13}), can be shown to be a direct consequence of applying the theory of ensembles on configuration space to classical general relativity. By contrast, the derivation of the rate equations from the Hamilton-Jacobi formalism alone \cite{G69} requires a much more lengthy derivation.

\section{Ground state Gaussian functional solution}\label{CSGFS}

Consider the Schr\"{o}dinger functional equation
\begin{equation}
-\frac{1}{2\Lambda R^2}
\left(\frac{\delta ^{2}A}{\delta \phi ^{2}}\right)+ \left[ \lambda\frac{\Lambda R^2}{2} + V + \frac{R^2 }{2 \Lambda}\phi ^{\prime 2} + \frac{\Lambda R^{2} m^2}{2 } \phi ^{ 2} \right] A = 0.
\end{equation}
with the ansatz
\begin{equation}
A \sim \exp\left\{ -\frac{1}{2} \int \int dy \, dz \, \Lambda_y \, \Lambda_z \, R_y^2 \, R_z^2 \; \phi_y \, K_{yz} \, \phi_z \right\}.
\end{equation}
If I calculate $\frac{\delta ^{2}A}{\delta \phi ^{2}}$ and collect terms that have the same powers of $\phi $, I get the following two equations for the kernel $K_{xy}$,
\begin{equation}\label{termsphi0}
\int dr\,
N \left[ \lambda\frac{\Lambda R^2}{2} + V(R,\Lambda) + \frac{\Lambda_r R_r^2}{2} K_{rr}
\right]=0
\end{equation}
and
\begin{equation}\label{termsphi2}
\int dr\, \Lambda_r R_r^2 \, N \left[
\frac{\phi_r^{\prime 2}}{2 \Lambda_r^2}  + \frac{m^2}{2}\phi_r^{ 2}
- \frac{1}{2} \int \int dydz \,\Lambda_y R_y^2 \Lambda_z R_z^2 \,\phi _{y}K_{yr}K_{rz}\phi _{z}
\right]=0 .
\end{equation}
After an integration by parts in Eq. (\ref{termsphi2}), the equations for $K_{xy}$ are of a type that is standard in the context of the Schr\"{o}dinger functional representation of quantum field theory in curved spacetimes \cite{K04,J90,H92,LS98}.

I want a solution that is valid for a foliation of spaces of constant positive curvature and a lapse function $N$ that is constant. Note that the $\Lambda$ and $R$ that appear in the line element of Eq. (\ref{gssg}) satisfy $\sqrt{h}=\Lambda R^2$ and $h^{rr}=\Lambda^{-2}$, where $h^{kl}$ is the inverse metric tensor on the three-dimensional spatial hypersurface of constant curvature. Then, Eq. (\ref{termsphi2}) can be written in the form
\begin{equation}\label{termsphi2v2}
\int dr \sqrt{h_r} \left[ \frac{1}{2}h^{rr}\frac{\partial \phi_r}{\partial r}\frac{\partial \phi_r}{\partial r} + \frac{m^2}{2}\phi_r^{ 2}
- \frac{1}{2} \int \int dy dz \sqrt{h_y} \sqrt{h_z}\,\phi _{y}K_{yr}K_{rz}\phi _{z} \right] = 0,
\end{equation}
and one can show that $K_{xy}$ satisfies
\begin{equation}\label{termsphi2v3}
\int dr \sqrt{h_r} K_{yr}K_{rz} = \left[ -\frac{1}{\sqrt{h_y}}\frac{\partial}{\partial y}\left( h^{yy} \sqrt{h_y} \frac{\partial}{\partial y} \right) + m^2 \right] \delta(y,z)
\end{equation}
where $\delta(y,z)=\frac{1}{\sqrt{h_y}}\delta(y-z)$ is the delta function on the hypersurface.

To get an explicit expression for $K_{xy}$ that solves Eq. (\ref{termsphi2v3}), introduce a fixed, particular set of coordinates for the line element of Eq.~(\ref{gssg}). Let
\begin{equation} \label{cfu}
N = 1, ~~~~~ N^r=0, ~~~~~ \Lambda=a_0, ~~~~~ R=a_0 \, \sin \, r,
\end{equation}
where $r\in \lbrack 0,2\pi )$ and $a_0$ can be interpreted as the scale factor of a closed Robertson-Walker universe. Then the solution of Eq. (\ref{termsphi2v3}) is given by
\begin{equation}
K_{xy} = \frac{1}{2 a_0^4} \sum_n \sqrt{\gamma_n } \, \psi^{(n)}_x \psi^{(n)}_y, \nonumber
\end{equation}
where the basis functions $\psi^{(n)}_r$ are solutions of a Schr\"{o}dinger-type equation in a space of constant curvature,
\begin{equation}
-\frac{1}{\sin^2r}\,\frac{\partial}{\partial r} \left( \sin^2 r \, \frac{\partial \psi_r^{(n)}}{\partial r} \right) +  m^2 a_0^2 \, \psi_r^{(n)} = \gamma_n \psi_r^{(n)}. \nonumber
\end{equation}
The $\psi_n(r)$ satisfy orthonormality and completeness relations. The eigenvalues $\gamma_n$ are given by
\begin{equation}
\gamma_n = n^2 - 1 + m^2 a_0^2,~~~~~n=1,2,3...~~. \nonumber
\end{equation}

Given the solution of Eq.~(\ref{termsphi2}), one can use Eq.~(\ref{termsphi0}) to express $a_0$ in terms of the cosmological constant and the energy $E$ of the quantized scalar field, since \cite{LS98}
\begin{equation}
E \sim a_0^3 \int dr\, \sin^2r \, K_{rr}. \nonumber
\end{equation}
However, $\int dr \, \sin^2r \, K_{rr} \sim \sum_{n} \gamma_n$, which diverges. This is a consequence of the infinite zero-point energy of the quantum field. Therefore, to extract a finite result for $a_0$ it becomes necessary to introduce renormalization procedures.

\end{document}